\documentclass[pre, twocolumn,  superscriptaddress, nopacs, nofootinbib]{revtex4}
\usepackage{bm}
\usepackage{epsfig}
\usepackage{amsmath}
\usepackage{amsfonts}
\usepackage{longtable}
\usepackage{epstopdf}
\setlongtables
\begin{document}

\title{Corresponding States of Structural Glass Formers}
\author{Yael S. Elmatad}
\author{David Chandler}
\email{chandler@cchem.berkeley.edu}
\affiliation{Department of Chemistry, University of California, Berkeley, CA 94720, USA}
\author{Juan P. Garrahan}
\affiliation{Department of Physics and Astronomy, University of Nottingham, Nottingham, NG7 2RD, UK}
\date{\today}

\begin{abstract}
The variation with respect to temperature $T$ of transport properties of 58 fragile structural glass forming liquids (67 data sets in total) are analyzed and shown to exhibit a remarkable degree of universality. In particular, super-Arrhenius behaviors of all super-cooled liquids appear to collapse to one parabola for which there is no singular behavior at any finite temperature. This behavior is bounded by an onset temperature $T_{\mathrm{o}}$ above which liquid transport has a much weaker temperature dependence. A similar collapse is also demonstrated, over the smaller available range, for existing numerical simulation data. 
\end{abstract}

\maketitle

Figure 1 shows the collapse of transport data for fragile glass-forming liquids.  These refer to super-cooled liquids where the increase of relaxation time $\tau$ with decreasing temperature $T$ is more rapid that that of the Arrhenius law, $\log(\tau / \tau_{\mathrm{R}}) = E\,\,(1/T - 1/T_{\mathrm{R}})$.  Here, $\tau_{\mathrm{R}}$ denotes a relaxation time at a reference temperature $T_{\mathrm{R}}$, and $E$ stands for activation energy over Boltzmann's constant $k_\mathrm{B}$.   Rather than linear in $1/T$, the collapsed data for $\log\tau$ is quadratic in $1/T$.  The data provide no evidence for singular expressions like the Vogel-Fulcher-Tammann (VFT) $\log\tau\sim\mathrm{const}/(T-T_{\mathrm{K}})$ or the mode-coupling $\log\tau\sim\mathrm{const}|\log(T-T_{\mathrm{c}})|$, where $T_{\mathrm{K}}$ or $T_{\mathrm{c}}$ are finite positive temperatures.  These forms are often used to fit transport data of super-cooled liquids \cite{Ediger_JPhysChem_1996}, and theoretical arguments have been presented as derivations of these forms  \cite{Adam_JChemPhys_July_1965_Xia_PhysRevLett_Jun_2001}.  

This relaxation time $\tau$ is an ``equilibrium'' property.  Its value is determined by the thermodynamic state of the system, and nothing more.  The rate of preparation, for example, is not pertinent. In contrast, one could consider transport properties when, for example, cooling or warming rates exceed relaxation rates, or at other irreversible conditions like those with which glass is manufactured.  In irreversible or driven cases, transport properties can be singular \cite{S_Ensemble}.   But in this article, we confine our attention to the reversible case, because it is this case where the great majority of quantitative measurements have been made.  It is for this case that we are showing with Figure 1 that there is no evidence of singular behavior controlled by the thermodynamic variable $T$.

\begin{figure*}[t!] 
   \centering
   \includegraphics[width = 6.4in]{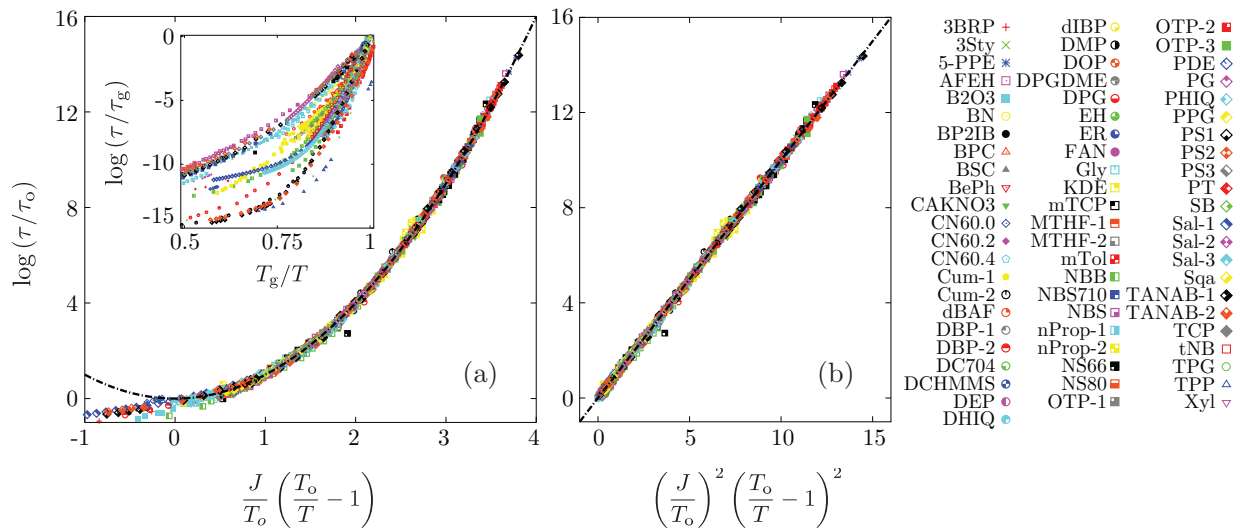} 
   \caption{(a) Collapse to a parabolic form of the structural relaxation times, $\tau$, and viscosities, $\eta$, as functions of temperature $T$ for fragile glass forming liquids.  Parameters $\tau_{\mathrm{o}}$, $T_{\mathrm{o}}$ and $J$ are listed in Table~I. Inset shows the same data when graphed in Angell-type plots, where $T_{\mathrm{g}}$ refers to the temperature at which the viscosity of the liquid is $10^{13}$ Poise or when the relaxation time reaches $10^2$ seconds.   (b) Data for temperatures $T<T_{\mathrm{o}}$ graphed as a function of the square of the collapse variable.  Key at right lists the 67 liquid data sets considered in the graphs.  The meaning of each acronym is given in Table~I.}
   \label{fig:fig_1}
\end{figure*}

Recently, Dyre and co-workers \cite{Hecksher_NatPhys_July_2008} arrived at a similar conclusion:  existing transport data support neither the idea of a finite temperature divergence nor the VFT formula.  All the data considered in Ref.\ \cite{Hecksher_NatPhys_July_2008} and more are treated here.  Our analysis takes the further step of collapsing the data and thus demonstrating universality.  A similar collapse was noted a few years ago, when R\"ossler and co-workers showed how seemingly varied behaviors for the transport properties of several super-cooled liquids could be represented by a single function of temperature \cite{Rossler_JNonCrysSol_1998}.  The fitting in Figure 1 differs mainly in the functional form adopted for the data collapse.  This difference enables a universal fit over a range of temperatures larger than those of Ref.\ \cite{Rossler_JNonCrysSol_1998}.    Kivelson and co-workers~\cite{Kivelson} have proposed another collapse to a non-singular function, but one with one more adjustable parameter than we consider.

We have used the quadratic form in earlier work \cite{Garrahan_PNAS_Aug_2003}. This form could be interpreted in terms of a random energy model, where activation barriers are assumed to be distributed as Gaussian variables \cite{Bassler_RhysRevLet_1987}, but the origin of this distribution would then remain to be explained.  We arrived at the quadratic form differently, from a class of kinetically constrained models \cite{KCM} where the activation energy to relax a domain of length scale $\xi$ grows as $\log\xi$ \cite{East}.  The equilibrium domain size is $\xi \simeq c^{-1/d}$, where $c$ is an equilibrium concentration of localized excitations and $d$ is dimensionality.  The Boltzmann distribution gives $\log c \propto 1/T$, from which one predicts an activation energy that grows as $1/T$ so that $\log\tau$ grows as $1/T^2$ (to leading order in $1/T$)  \cite{East}.

We expect the range of validity of this quadratic behavior to be bounded~\cite{Garrahan_PNAS_Aug_2003}.   In particular, it should not apply above a temperature $T_{\mathrm{o}}$ where excitations facilitating molecular motions are present throughout the system.  In that regime, correlated dynamics is not required for molecular motions, and accordingly, temperature variation of transport is nearly negligible \cite{Jonas_Science_Jun_1982_Chandler_Science_1983}.  The quadratic form should also not apply below another temperature, which we call $T_{\mathrm{x}}$. The reasoning here recognizes that correlated dynamics leading to super-Arrhenius behavior \cite{Palmer_PRL_1984} is the result of constraints due to intermolecular forces.   At an energetic cost, $E$, these constraints can be avoided.  The time scale to pay that cost is $\tau_{\mathrm{x}} \exp(E/T)$.  While this time can be very long, at a low enough temperature it will become shorter than a super-Arrhenius time.  This is the temperature  $T_{\mathrm{x}}$, below which relaxation will be dominated by dynamics that avoid constraints.  Therefore, we use
\begin{equation}
\log(\tau / \tau_{\mathrm{o}})  \simeq  (J/T_{\mathrm{o}})^2 (T_{\mathrm{o}} / T - 1 )^2, \,\, \,\,\,T_{\mathrm{o}} > T > T_{\mathrm{x}} \,,
\end{equation} 
to fit data in Figure 1, using $J$ as the parameter to set the energy scale for excitations of correlated dynamics, and with the understanding that for $T>T_{\mathrm{o}}$, $\log(\tau / \tau_{\mathrm{o}})$ has little temperature dependence, and for  $T<T_{\mathrm{x}}$, $\log(\tau / \tau_{\mathrm{o}})$ will crossover to Arrhenius temperature dependence.

The onset and crossover temperatures are material properties that may or may not fall within the range of specific experiments.  For systems where $T_{\mathrm{o}}$ approaches $T_{\mathrm{x}}$, fragile behavior will not be observed.  Most data that we have found lies in either the fragile and normal liquid regime, $T>T_{\mathrm{x}}$, or in the strong regime, $T<T_{\mathrm{x}}$, but not both.  In Ref.~\cite{Garrahan_PNAS_Aug_2003}, we noted published transport data on two organic liquids that appear to exhibit the crossover~\cite{Laughlin_JPhysChem_1972}.  But for one of these, salol, other data seem to contradict this finding~\cite{Dixon_PhysRevE_Aug_1994, Richert_JChemPhys_June_1998}.  Yet a change from super-Arrhenius to Arrhenius behavior should also be reflected in a change of transport de-coupling~\cite{Pan_ChemPhysChem_2005}, and this change is seen between dielectric and viscous relaxation~\cite{Chang_JPhysChemB_1997}.  The temperature dependence of transport~\cite{Liu_PRL_2005} and of decoupling~\cite{Chen_PNAS_June_2006} in films of liquid water also suggest the presence of a crossover.  Unfortunately, the amount of data available at present is too sparse to make a convincing case for the origin of this phenomenon.  As such, for the present, we focus on the fragile regime.

Table~I collects the parameters obtained in the fitting data to Eq.(1) with $T_{\mathrm{o}}>T>T_{\mathrm{x}}$.  For each liquid considered, the data set for this regime contained five or more data points, and most contained ten or more data points.   This is the data shown in Figure 1.  Some of the data refer to viscosity measurements, others refer to relaxation time measurements.  We use the same formula for both, replacing $\tau$ with $\eta$ when referring to viscosity.  For each liquid, the three fitting parameters are determined by minimizing the mean square deviation, $\sigma^2$, between the data and the quadratic form for temperatures that exceed a preliminary estimate of the onset temperature.  This estimate is the highest temperature at which the curvature of the data, as a function of $1/T$, appears to take on its maximum value.  This estimate usually coincides closely with the value of $T_\mathrm{o}$ found from fitting the quadratic form.  The standard deviations obtained by this fitting are noted in Table~I.  
Also noted in Table~I are the standard deviations obtained by fitting the VFT form to the same data.  Both the quadratic form and the VFT form have three independent parameters. Considering all 67 liquid data sets, the mean standard deviation for the parabolic form is $0.073 \pm 0.073$ and for the VFT form is $0.088 \pm 0.14$.  While the standard deviations are similar, there are at least two reasons to favor the quadratic form over the VFT form.  The first \cite{Hecksher_NatPhys_July_2008} is that the quadratic form does not require the introduction of a metaphysical Kauzmann temperature -- an implausible thermodynamic state point that by definition is unobservable \cite{Stillinger_JChemPhys_Mar_1988}.   The second is that,  for approximately half the liquids fitted, the VFT form achieves small standard deviations with a pre-factor time that is less than 10 fs, which is too short to coincide with structural relaxation at any reference state of a molecular liquid.    

The reference time, $\tau_{\mathrm{o}}$, is the time for relaxing a microscopic region of liquid at the onset temperature.  We expect these times to be significantly larger than 1 ps.  Similarly, we expect the reference viscosity, $\eta_{\mathrm{o}}$, to be not much smaller than 1 Poise.   A much smaller value can be an indication of treating a strong material as if it were fragile.  For example, fitting available data for liquid 3-phenyl-1-propanol (3Ph1P) \cite{Hecksher_NatPhys_July_2008, Igarashi_RevSciInst_2008} with Eq.(1) yields a seemingly acceptable standard deviation of $\sigma=0.16$, but the energy scale compared to the reference temperature is curiously low, $J/T_{\mathrm{o}}\approx1.2$, and the reference time is unreasonably short, $\tau_{\mathrm{o}}\approx10^{-16}$ s.  Instead, by fitting to the Arrhenius form with $\log \left(\tau_\mathrm{R}/\mathrm{s} \right) = -2.4$ at $T=T_\mathrm{R} = 200 \mathrm{K}$, the activation energy $E$ and standard deviation $\sigma$ have reasonable values of $40T_\mathrm{R}$ and 0.57, respectively. Another similar case is the liquid triphenyl-ethylene (TPE) \cite{Jakobsen_JChemPhys_Dec_2005}.  Again, while the standard deviation $\sigma = 0.0086$ is small, the time is unreasonably short, $\tau_{\mathrm{o}}\approx10^{-14}$ s.  An Arrhenius fit for these data yields $\log \left(\tau_\mathrm{R}/\mathrm{s} \right) = -3.1$, $T_\mathrm{R} = 274 \mathrm{K}$, $E/T_\mathrm{R} = 58$ and $\sigma=0.058$.  The available data therefore suggests that these super-cooled liquids are strong.  That is, the crossover temperature is larger than any temperature for which the data is available: $T_{\mathrm{x}} > T_R$.  Perhaps for these liquids, or others like them, relaxation could be studied at higher temperatures to find evidence for a crossover temperature.

There is one outlying data point on the graphs of Figure~1  for so called NS 66 \cite{Neuville_ChemGeo_2006}.  This occurs at a state point far separated from all the other state points for which other data points exist.  We suspect this point might be erroneous.  

\begin{figure}[t!] 
   \centering
   \includegraphics{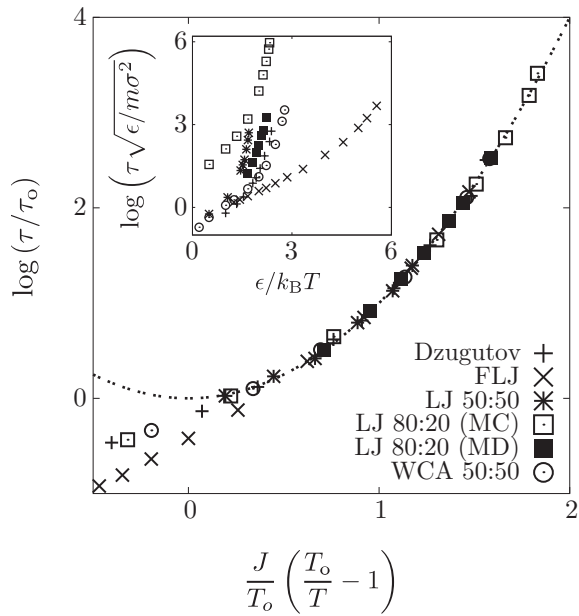} 
 \caption{Collapse to a parabolic form of the structural relaxation times, $\tau$, as functions of temperature $T$ for simulations of models of fragile glass forming liquids.  Parameters $\tau_{\mathrm{o}}$, $T_{\mathrm{o}}$ and $J$ are listed in Table~II. Inset shows the same data when graphed as $\log \tau$ vs $1/T$.  $T$ is given in units of $\epsilon/k_\mathrm{B}$ and $\tau$ in units of $\sqrt{m \sigma^2/\epsilon}$.  Here, $m$ is a particle mass, $\sigma$ is a particle diameter, and $\epsilon$ is an energy parameter that characterizes interparticle interactions. See Refs. \cite{Gebremichael_PhdThesis_2004, Shintani_NatMat_Oct_2008, Lacevic_JChemPhys_2003, Berthier_JPhysCondensMatt_2007, Karmakar_CondMat_2008, Hedges_JChemPhys_2007} for the precise meaning in each particular case.   The meaning of each acronym is given in Table~II.}
   \label{fig:fig_2_v_1}
\end{figure}

Figure~2 shows quadratic data collapse for data from six numerical simulations of fragile glassfomers.  Here, the fitting was done using the same methods used to collapse the data in Figure~1(a).  Table~II shows the parameters used for the data collapse. $J/T_\mathrm{o}$ for the numerical simulation data is comparable in magnitude to the values obtained for many of the experimental liquids.  
The simulation data extends over three or four orders of magnitude while the experimental data extends over more than ten orders of magnitude.
\\
\\

\begin{acknowledgments}
A number of scientists generously provided data sets that greatly assisted us in this work: 
L.~Berthier, 
S.-H.~Chen, 
J.~Dyre, 
Y.~Gebremichael, 
T.~Hecksher, 
S.~Karmakar, 
A.~S.~Keys,
 L.~Maibaum,
S.~Nagel, 
R.~Richert, 
S.~Sastry,
and H. Tanaka.
Y.~S.~E. was supported by NSF GRFP and ONL NDSEG fellowships. D.~C. was supported by NSF. J.~P.~G. was supported by EPSRC grant no.\ GR/S54074/01.
D.~C. was Overseas Visiting Scholar of St. John's College, Cambridge during the time that this paper was written.

\end{acknowledgments}

\begin{center}
\renewcommand{\thefootnote}{\alph{footnote}}
\begingroup
\begin{longtable*}{c c c c c c c c c c c}
\caption{Fragile Glass Formers}\\
DCHMMS &  ether-2-ethyl-hexylamine & 1000 & 0.00 & 0.00 & 0.00 & 0.0000 (0.0000) & 1000 & 1000 & 1000-1000 & [1, 1] \kill
\hline\hline \\[-2ex]
    \multicolumn{1}{c}{System\ } &
  \multicolumn{1}{c}{Full Name \ } &
  \multicolumn{1}{c}{$T_\mathrm{o}\mathrm{/K}\tablenotemark[1] \ $} &
  \multicolumn{1}{c}{$J/T_\mathrm{o}\tablenotemark[2] \ $} &
   \multicolumn{1}{c}{$\log \tau_\mathrm{o}\mathrm{/s}\tablenotemark[3] \ $} &
   \multicolumn{1}{c}{$\log \eta_\mathrm{o}\mathrm{/P}\tablenotemark[4] \ $} &
   \multicolumn{1}{c}{$\sigma \ (\sigma_\mathrm{VFT})\tablenotemark[5]$ \ } &
   \multicolumn{1}{c}{$T_\mathrm{m}\mathrm{/K}\tablenotemark[6] \ $} &
   \multicolumn{1}{c}{$T_\mathrm{g}\mathrm{/K}\tablenotemark[7] \ $} &
   \multicolumn{1}{c}{Range/K\tablenotemark[8] \ } &
\multicolumn{1}{c}{Ref.} \\[0.5ex] \hline
   \\[-1.8ex]
\endfirsthead

\multicolumn{11}{c}{{\tablename} \thetable{} -- Continued} \\[0.5ex]
  \hline \hline \\[-2ex]
    \multicolumn{1}{c}{System\ } &
  \multicolumn{1}{c}{Full Name \ } &
  \multicolumn{1}{c}{$T_\mathrm{o}\mathrm{/K}\tablenotemark[1] \ $} &
  \multicolumn{1}{c}{$J/T_\mathrm{o}\tablenotemark[2] \ $} &
   \multicolumn{1}{c}{$\log \tau_\mathrm{o}\mathrm{/s}\tablenotemark[3] \ $} &
   \multicolumn{1}{c}{$\log \eta_\mathrm{o}\mathrm{/P}\tablenotemark[4] \ $} &
   \multicolumn{1}{c}{$\sigma \ (\sigma_\mathrm{VFT})\tablenotemark[5]$ \ } &
   \multicolumn{1}{c}{$T_\mathrm{m}\mathrm{/K}\tablenotemark[6] \ $} &
   \multicolumn{1}{c}{$T_\mathrm{g}\mathrm{/K}\tablenotemark[7] \ $} &
   \multicolumn{1}{c}{Range/K\tablenotemark[8] \ } &
\multicolumn{1}{c}{Ref.} \\[0.5ex] \hline
  \\[-1.8ex]
\endhead

\\[-1.8ex] \hline \hline
\endfoot

\\[-1.8ex] \hline \hline
\multicolumn{11}{l}{ \footnotemark[1]{$T_\mathrm{o}$ is the fitted onset temperature in K.} } \\
\multicolumn{11}{l}{\footnotemark[2]{$J$ is the fitted energy scale over $k_\mathrm{B}$.}} \\ 
\multicolumn{11}{l}{\footnotemark[3]{$\tau_\mathrm{o}$ is the fitted onset relaxation time in seconds.} }\\
\multicolumn{11}{l}{\footnotemark[4]{$\eta_\mathrm{o}$ is the fitted onset viscosity in Poise.}} \\
\multicolumn{11}{l}{\footnotemark[5]{$\sigma$ is the standard deviation of the quadratic form given by: $\left ({{1}/({N-n}) \sum_i \left ( \log_{10} \tau_{\mathrm{fit},i}  - \log_{10} \tau_{\mathrm{data},i} \right)^2} \right )^{1/2}$. $N$ is the number }}\\ 
\multicolumn{11}{l}{ \ of fitted data points, $n=3$ is the number of degrees of freedom for all reported fits. $i = \{ 1, N \}$ indexes the fitted points.} \\ 
\multicolumn{11}{l}{ \ $\sigma_\mathrm{VFT}$ is the standard deviation for fitting the parameters $\tau^\mathrm{VFT}_\mathrm{o}, A$ and $T_\mathrm{K}$ of the VFT form: $\tau = \tau_\mathrm{o}^\mathrm{VFT} \exp\left(A/\left(T-T_\mathrm{K}\right)\right)$.} \\
\multicolumn{11}{l}{\footnotemark[6]{$T_\mathrm{m}$ is the melting temperature.}} \\ 
\multicolumn{11}{l}{\footnotemark[7]{$T_\mathrm{g}$ is the glass transition temperature i.e., where $\eta = 10^{13}$ P or $\tau  = 10^2$ s.}} \\
\multicolumn{11}{l}{\footnotemark[8]{The range of temperature for data reported in K.  Only data for $T < T_\mathrm{o}$ is fitted.}} \\

\endlastfoot
3BRP	&	3-bromopentane	&	192	&	4.3	&	-9.4	&		& 0.13 (0.13) &	147	&	108	&	107-289	&	\cite{Berberian_JChemPhys_June_1986}	\\
3Sty	&	3-styrene	&	314	&	8.5	&	-6.5	&					& 0.024 (0.025) &	242	&	237	&	235-280	&	\cite{Blockhowicz_PhdThesis_2003}	\\
5-PPE	&	5-polyphenyl-ether	&	398	&	6.2	&	-12.4	& 	&  0.0044 (0.058)  &		&	248	&	248-264	&		\cite{Jakobsen_JChemPhys_Dec_2005}	\\
AFEH	&	2-phenyl-5-acetomethyl-5-	&	285	&	9.4	&	-6.1	& 	& 0.0038 (0.0096)  &		&	219	&	220-240	&	\cite{Hecksher_NatPhys_July_2008}	\\
		&	ethyl-1,3-dioxocylohexane	& \\
B2O3	&	boron oxide (B$_2$O$_3$)	&	1066 	&	3.3	&		&	3.0	&	0.095 (1.0) & 723	&	541	&	533-1665	&	\cite{Tweer_JChemPhys_March_1971}	\\
BePh	&	benzophenone	&	328	&	6.3	&	-11.0	&	&		0.05 (0.052) &	321	&	208	&	215-240	&	\cite{Lunkenheimer_PhysRevE_March_2008}	\\
BN	&	butyronitrile	&	135	&	6.6	&	-4.8	&		&		0.025 (0.021) &116	&	97	&	97-116	&	\cite{Ito_JChemPhys_July_2006}	\\
BP2IB	&	biphenyl-2yl-isobutylate	&	313	&	6.7	&	-9.4	&&	0.0063 (0.0084) &	&	209	&	210-232	&	\cite{Hecksher_NatPhys_July_2008}	\\
BPC	&	3,3,4,4,-benzphenonetetra-	&	432	&	9.9	&	-6.8	&		&	0.014 (0.007) &	&	333	&	334-362	&	\cite{Ngai_JChemPhys_2004}	\\
	&	carboxylicdianhydride \\
BSC	&	Borosilicate Crown glass	&	2002	&	2.3	&		&	1.9 	&0.075 (0.15) &		&	825	&	800-1594	&	\cite{Tweer_JChemPhys_March_1971}	\\
CaKNO3	&	Ca-K-NO$_3^-$	&	444	&	10.8	&		&	0.3	&0.37 (0.22)&		&	338	&	341-668	&	\cite{Angell_JNonCrysSol_1988}	\\
CN60.0	&	soda lime silicate glass.0	&	1702	&	5.2	&		&	1.2	&	 0.046 (0.061) &	&	1030	&	1012-1809	&	\cite{Neuville_ChemGeo_2006}	\\
CN60.2	&	soda lime silicate glass.2	&	1668	&	3.2	&		&	1.8	&	0.086 (0.019) &	&	820	&	803-1563	&	\cite{Neuville_ChemGeo_2006}	\\
CN60.4	&	soda lime silicate glass.4	&	1929	&	1.9	&		&	1.6	&	0.19 (0.045) &	&	700	&	684-1563	&	\cite{Neuville_ChemGeo_2006}	\\
Cum-1	&	isopropyl-benzene	&	174	&	8.6	&	-6.9	&		&0.028 (0.018) &	177	&	129	&	130-149	&	\cite{Hecksher_NatPhys_July_2008}	\\
Cum-2	&	isopropylbenzene	&	194	&	6.8	&		&	-0.6	&0.21 (0.32) &	177	&	129	&	129-306	&	\cite{Tweer_JChemPhys_March_1971}	\\
dBAF	&	dibutyl-ammonium formate	&	220	&	6.6	&	-5.9	&	&	0.097 (0.031) 	&		&	155	&	156-200	&	\cite{Ito_JPhysChemB_2006} \\	
DBP-1 &	dibutyl-phtalate	&	241	&	8.3	&	-6.2	&		& 0.052 (0.026) &	238  & 	179	&	180-224	&	\cite{Olsen_PhysRevLeett_Feb_2001}	\\
DBP-2	&	di-n-butylphtalate	&	320	&	4.1	&		&	-0.9	&	0.2 (0.36) & 	&	168	&	178-369	&	\cite{Tweer_JChemPhys_March_1971}	\\
DC704	&	tetraphenyl-tetramethyl-	&	306	&	7.9	&	-9.8	&	  & 0.0097 (0.019)	&		&	213	&	211-240	&	\cite{Jakobsen_JChemPhys_Dec_2005}	\\
		& trisiloxane \\
DCHMMS	&	dichyclohexyl-methyl-	&	275	&	10.9	&	-5.4	&	& 0.0072 (0.0089)	&		&	221	&	220-240	&	\cite{Diaz-Calleja_PhysRevE_2005}	\\
		& 2-methylsuccinate\\
DEP	&	diethyl-phtalate	&	262	&	7.3	&	-7.6	&		& 0.024 (0.0098) &	270	&	185	&	186-222	&	\cite{Hecksher_NatPhys_July_2008,Igarashi_RevSciInst_2008}	\\
DHIQ	&	decahydroisoquinoline	&	197	&	25.8	&	-4.3	&		&	0.042 (0.077) &	&	180	&	180-192	&	\cite{Jakobsen_JChemPhys_Dec_2005}	\\
dIBP	&	di-iso-butyl-phtalate	&	247	&	9.7	&	-5.4	&		&	0.0028 (0.02)	& &	194	&	195-221	&	\cite{Hecksher_NatPhys_July_2008}\\
DMP	&	dimethyl-phtalate	&	261	&	8.5	&	-6.4	&		&0.017 (0.0077) &	275	&	195	&	196-220	&	\cite{Hecksher_NatPhys_July_2008,Igarashi_RevSciInst_2008}
	\\
DOP	&	dioctyl-phtalate	&	251	&	7.8	&	-5.1	&		&0.023 (0.0036) &	223	&	187	&	188-220	&	\cite{Hecksher_NatPhys_July_2008}	\\
DPG	&	dipropylene-glycol	&	268	&	7.8	&	-5.9	&		&0.043 (0.021) &	$<$234	&	196	&	196-240	&	\cite{Hecksher_NatPhys_July_2008, Igarashi_RevSciInst_2008}	\\
DPGDME	&	dipropyle-glycol-	&	177	&	9.9	&	-6.0	&		& 0.015 (0.022) &		&	136	&	139-155	&	\cite{Hecksher_NatPhys_July_2008}	\\
		& dimethyl-ether \\
EH	&	ether-2-ethyl-hexylamine	&	183	&	9.1	&	-5.6	&		& 0.028 (0.011)	 &197	&	140	&	142-166	&	\cite{Wang_JPhysChemB_2005}	\\
ER	&	diglycidyl-ether-of- 	&	309	&	14.0	&	-6.7	&		& 0.014 (0.035)&	325	&	255	&	259-291	&	\cite{Mierzwa_JChemPhys_2008}	\\
	&	bisphenol A (epoxy-resin)\\
FAN	&	3-fluoro-aniline	&	225	&	10.3	&	-7.2	&		&	0.16 (0.16)	 &	&	173	&	173-198	&	\cite{Widersich_JPhysCondensMat_March_1999}	\\
Gly	&	glycerol	&	338	&	4.1	&	-7.7	&		& 0.033 (0.0053)	 &	293	&	191	&	192-252	&		\cite{Olsen_PhysRevLeett_Feb_2001}\\
KDE	&	cresolphthalein-	&	461	&	7.1	&	-8.0	&		&0.0095 (0.025)	 &	387	&	318	&	315-383	&	\cite{Paluch_JChemPhys_2001}	\\
	& dimethylether \\
mTCP	&	$m$-tricresyl-phosphate	&	270	&	9.2	&	-5.6	&		&0.018 (0.0095)	 &	299	&	208	&	209-233	&	\cite{Blochowicz_JChemPhys_Apr_2006}	\\
MTHF-1	&	2-methyltetrahydrofuran	&	119	&	9.8	&	-6.9	&		& 0.022 (0.026)	 &	137	&	92	&	91-108	&	\cite{Hecksher_NatPhys_July_2008}	\\
MTHF-2	&	2-methyltetrahydrofuran	&	126	&	8.5	&	-8.8	&		& 0.049 (0.11) &	137	&	91	&	94-179	&	\cite{Richert_JChemPhys_June_1998} 	\\
mTol	&	$m$-toluene	&	237	&	10.6	&	-6.6	&		& 0.01 (0.0051)	 &		&	185	&	184-200	&	\cite{Hecksher_NatPhys_July_2008, Igarashi_RevSciInst_2008}
	\\
NBB	&	$n$-butylbenzene	&	202	&	5.9	&		& 	2.0	& 0.25 (0.16)	 &	185	&	129	&	135-306	&	\cite{Tweer_JChemPhys_March_1971}	\\
NBS	&	NBS-711 standard	&	2780	&	1.1	&		&	1.5	& 0.1 (0.062)	 &		&	705	&	665-1614	&	\cite{Tweer_JChemPhys_March_1971}	\\
NBS 710	&		&	2483	&	1.7	&		&	1.5	& 0.097 (0.018)	 &		&	830	&	827-1776	&	\cite{Neuville_ChemGeo_2006}	\\
nProp-1	&	$n$-propanol	&	350	&	1.4	&	-10.7	&		& 0.12 (0.17)	 &	147	&	99	&	100-300	&	\cite{Richert_JChemPhys_June_1998} 	\\
nProp-2	&	$n$-propanol	&	398	&	1.2	&		&	-2	& 0.21 (0.27) &	147	&	99	&	104-370	&	\cite{Tweer_JChemPhys_March_1971}	\\
NS 66	&		&	2489	&	1.4	&		&	1.2	&	0.28 (0.19)	&	 &	726	&	719-1805	&	\cite{Neuville_ChemGeo_2006}	\\
NS 80	&		&	2435	&	1.5	&		&	1.7	&	0.10 (0.065)	&	 &	758	&	718-1759	&	\cite{Neuville_ChemGeo_2006}	\\
OTP-1	&	$o$-terphenyl 	&	341	&	8.5	&	-8.9	&		&0.038 (0.035)	 &	329	&	243	&	252-282	&	\cite{Richert_JChemPhys_Oct_2005}	\\
OTP-2	&	$o$-terphenyl 	&	340	&	8.6	&		&	0.0	& 0.066 (0.064)	 &	329	&	240	&	239-267	&	\cite{Laughlin_JPhysChem_1972}	\\
OTP-3	&	$o$-terphenyl 	&	357	&	7.7	&	-9.9	&		& 0.16 (0.18)	 &	329	&	246	&	248-311	&	\cite{Richert_JChemPhys_June_1998} 	\\
PDE	&	phenolphthalein-	&	397	&	9.3	&	-7.7	&	 &	0.022 (0.031)	&	373	&	294	&	299-333	&	\cite{Hensel-Bielowka_PhysRevLett_2002}	\\
	& 	dimethylether \\
PG	&	1,2-propandiol 	&	321	&	3.4	&	-7.7	&		&0.0062 (0.0046)	 &	214	&	164	&	180-211	&	\cite{Hecksher_NatPhys_July_2008}	\\
	& (propylene-glycol) \\
PHIQ	&	perhydroisoquinoline	&	208	&	18.5	&	-5.8	&		&0.14 (0.055)	 &		&	181	&	182-206	&	\cite{Hecksher_NatPhys_July_2008} \\
PPG	&	polypropylene-glycol	&	263	&	8.7	&	-6.1	&		&0.049 (0.0063)	 &	215	&	199	&	200-240	&	\cite{Olsen_PhysRevLeett_Feb_2001}	\\
PS1	&	titania-bearing sodium &	2395	&	1.7	&		&	1.5	& 0.078 (0.042)	 &		&	796	&	837-1591	&	\cite{Liska_ChemGeo_1996}	\\
 	& silicate melt \#1 \\
PS2	&	titania-bearing sodium	&	2688	&	1.3	&		&	1.3	& 0.074 (0.038)	 &		&	746	&	784-1679	&	\cite{Liska_ChemGeo_1996}	\\
 	& silicate melt \#2 \\
PS3	&	titania-bearing sodium 	&	2109	&	1.9	&		&	1.9	& 0.081 (0.06)	 &		&	765	&	815-1676	&	\cite{Liska_ChemGeo_1996}	\\
 	& silicate melt \#3 \\
PT	&	pyridine-toluene	&	146	&	17.5	&	-5.5	&		&0.019 (0.023)	 &		&	126	&	125-131	&	\cite{Olsen_PhysRevLeett_Feb_2001}	\\
Sal-1	&	salol	&	309	&	8.1	&	-8.5	&		& 0.05 (0.12)	 &	315	&	221	&	218-382	&	\cite{Dixon_PhysRevE_Aug_1994}	\\
Sal-2	&	salol	&	299	&	9.1	&	-8.3	&		&0.066 (0.048)	 &	315	&	222	&	223-253	&	\cite{Gainaru_PhysRevB_2005}	\\
Sal-3	&	salol	&	308	&	8.3	&	-8.5	&		& 0.069 (0.24)	 &	315	&	221	&	220-309	&	\cite{Richert_JChemPhys_June_1998} 	\\
SB	&	sucrose-benzonate	&	421	&	11.2	&	-5.8	&		&0.04 (0.019)	 &	373	&	340	&	341-400	&	\cite{Rajian_JChemPhys_2006}	\\
Sqa	&	squalane	&	224	&	8.3	&	-5.3	&		&0.092 (0.053)	 &	235	&	170	&	170-210	&	\cite{Jakobsen_JChemPhys_Dec_2005}	\\
TANAB-1	&	tri-$\alpha$-naphtylbenzene	&	519	&	6.8	&		&	-0.9	& 0.082 (0.34)	 &		&	335	&	332-584	&	\cite{Plazek_JChemPhys_1966}	\\
TANAB-2	&	tri-$\alpha$-naphtylbenzene	&	520	&	6.4	&		&	-0.8	& 0.1 (0.21) &		&	335	&	333-588	&	\cite{Tweer_JChemPhys_March_1971}	\\
TCP	&	tricresyl-phosphate	&	280	&	8.8	&	-6.3	&		&0.012 (0.013)	 &	240	&	209	&	216-248	&	\cite{Hecksher_NatPhys_July_2008, Igarashi_RevSciInst_2008}
	\\
tNB	&	trisnaphthylbenzene	&	510	&	7.1	&	-9.2	&		&	0.019 (0.023)	 &	& 	342	&	357-405	&	\cite{Richert_JChemPhys_Jan_2003}	\\
TPG	&	tripropylene-glycol	&	251	&	8.9	&	-5.5	&		& 0.041 (0.0055)	 &	232	&	192	&	192-228	&	\cite{Olsen_PhysRevLeett_Feb_2001}	\\
TPP	&	triphenyl phosphite	&	286	&	7.7	&		&	-0.5	& 0.08 (0.18)	 &	296	&	204	&	203-291	&	\cite{Martinez_Nature_Apr_2001}	\\
Xyl	&	xylitol	&	311	&	11.1	&	-5.8	&		& 0.026 (0.0057)	 &	367	&	250	&	254-284	&	\cite{Hecksher_NatPhys_July_2008}	\\ [1ex]
\hline
\label{table:fragile}
\end{longtable*}
\endgroup
\renewcommand{\thefootnote}{\arabic{footnote}}
\end{center}

\renewcommand{\thefootnote}{\alph{footnote}}
\begingroup
\begin{longtable*}{c c c c c c c c}
\caption{Fragile Glass Former Simulations}\\
\hline\hline \\[-2ex]

\multicolumn{1}{c}{System \ } &
  \multicolumn{1}{c}{Description \ } &
  \multicolumn{1}{c}{$T_o k_\mathrm{B}/\epsilon$\footnotemark[1] \ } &
  \multicolumn{1}{c}{$J/T_o\footnotemark[2] \ $} &
   \multicolumn{1}{c}{$\log \left(\tau_\mathrm{o}/ \sqrt{m \sigma^2/\epsilon} \right)\footnotemark[3] $} \ &
   \multicolumn{1}{c}{$\sigma$\footnotemark[4] \ } &
   \multicolumn{1}{c}{Range/($k_\mathrm{B}/\epsilon)\footnotemark[5] \ $} &
\multicolumn{1}{c}{Ref.} \kill

 \multicolumn{1}{c}{System \ } &
   \multicolumn{1}{c}{Description \ } &
  \multicolumn{1}{c}{$T_o k_\mathrm{B}/\epsilon$\footnotemark[1] \ } &
  \multicolumn{1}{c}{$J/T_o\footnotemark[2] \ $} &
   \multicolumn{1}{c}{$\log \left(\tau_\mathrm{o}/ \sqrt{m \sigma^2/\epsilon} \right)\footnotemark[3] $} \ &
   \multicolumn{1}{c}{$\sigma$\footnotemark[4] \ } &
   \multicolumn{1}{c}{Range/($k_\mathrm{B}/\epsilon)\footnotemark[5] \ $} &
\multicolumn{1}{c}{Ref.} \\[0.5ex] \hline

   \\[-1.8ex]
\endfirsthead

\multicolumn{8}{c}{{\tablename} \thetable{} -- Continued} \\[0.5ex]
  \hline \hline \\[-2ex]
   \multicolumn{1}{c}{System \ } &
  \multicolumn{1}{c}{Description \ } &
  \multicolumn{1}{c}{\footnotemark[1]$T_o k_\mathrm{B}/\epsilon$ \ } &
  \multicolumn{1}{c}{\footnotemark[2]$J/T_o \ $} &
   \multicolumn{1}{c}{\footnotemark[3]$\log \left(\tau_\mathrm{o}/ \sqrt{m \sigma^2/\epsilon} \right) $} \ &
   \multicolumn{1}{c}{\footnotemark[4]$\sigma$ \ } &
   \multicolumn{1}{c}{\footnotemark[5]Range/($k_\mathrm{B}/\epsilon) \ $} &
\multicolumn{1}{c}{Ref.} \\[0.5ex] \hline
  \\[-1.8ex]
\endhead

\\[-1.8ex] \hline \hline
\endfoot

\\[-1.8ex] \hline \hline
LJ 80:20 (MC) \ \ \ \ & Weeks-Chandler-Andersen \ \ \ & 0.00 & 0.00 & 0.00 & 0.0000  & 0.00-0.000 & [11] \kill
\multicolumn{8}{l}{ \footnotemark[1]{$T_\mathrm{o}$ is the fitted onset temperature in $k_\mathrm{B}/\epsilon$}} \\
\multicolumn{8}{l}{\footnotemark[2]{$J$ is the fitted energy scale over $k_\mathrm{B}$.}} \\ 
\multicolumn{8}{l}{\footnotemark[3]{$\tau_\mathrm{o}$ is the fitted onset relaxation time in $\sqrt{m \sigma^2/\epsilon}$.} }\\
\multicolumn{8}{l}{\footnotemark[4]{$\sigma$ is the standard deviation of the quadratic form given by: $\left ({{1}/({N-n}) \sum_i \left ( \log_{10} \tau_{\mathrm{fit},i}  - \log_{10} \tau_{\mathrm{data},i} \right)^2} \right )^{1/2}$. $N$ is the number }}\\ 
\multicolumn{8}{l}{ \ of fitted data points, $n=3$ is the number of degrees of freedom. $i = \{ 1, N \}$ indexes the fitted points.} \\ 
\multicolumn{8}{l}{\footnotemark[5]{The range of temperature for data reported in units of $k_\mathrm{B}/\epsilon$.  Only data for $T < T_\mathrm{o}$ is fitted.}} \\
\endlastfoot
Dzugutov	  &	Dzugatov 50:50 Mixture	&	0.8	&	1.8	&	0.3	&	 0.053 & 0.42-1 & \cite{Gebremichael_PhdThesis_2004} \\
FLJ  &	Frustrated Lennard-Jones	&	0.3	&	1.6	&	1.5	&	 0.0015  & 0.18-0.8 & \cite{Shintani_NatMat_Oct_2008} \\
LJ 50:50	 &	Lennard-Jones 50:50 Mixture	&	0.7	&	5.6	&	1.3	&	 0.027 & 0.59-2 & \cite{Lacevic_JChemPhys_2003} \\
LJ 80:20 (MC) 	&	Lennard-Jones 80:20 Mixture 	&	0.8	&	1.9	&	2.6	&	 0.054 & 0.43-2 & \cite{Berthier_JPhysCondensMatt_2007} \\
			& (Monte Carlo) \\
LJ 80:20 (MD) &	Lennard-Jones 80:20 Mixture 	&	0.8	&	1.9	&	0.7	&	 0.013 & 0.45-0.6 & \cite{Karmakar_CondMat_2008} \\
			& (Molecular Dynamics) \\
WCA 50:50 & Weeks-Chandler-Andersen	&	0.6	&	2.9	&	1	& 0.035 & 0.36-5 & \cite{Hedges_JChemPhys_2007} \\
 			&  50:50 Mixture & \\ [1ex]

\hline
\label{table:fragile}
\end{longtable*}
\endgroup
\renewcommand{\thefootnote}{\arabic{footnote}}


\bibliographystyle{./achemso}	

\pagebreak

\end{document}